\documentstyle[12pt,a4]{article}
\newcommand{\half}{\mbox{${\textstyle \frac{1}{2}}$}}           
\begin{document}
\baselineskip 4ex
\vspace*{-2cm}
\hfill{\bf TSL/ISV-96-0152}
\vspace{1.5cm}
\begin{center}
\vspace*{10mm}
{\Large{\bf Scattering  Wave Functions at Bound State Poles}}\\[7ex]
{\large G{\"{o}}ran F\"{a}ldt}\footnote{Electronic address: faldt@tsl.uu.se}
\\[1ex]
{\normalsize  Division of Nuclear Physics,  Box 533, 751 21 Uppsala, Sweden}
\\[3ex]
{\large Colin Wilkin}\footnote{Electronic address: cw@hep.ucl.ac.uk}\\[1ex]
{\normalsize  University College London, London, WC1E 6BT, UK}\\[3ex]
\today\\[4ex]
\end{center}

\begin{abstract}
The normalisation relation between the bound and scattering S-state wave
functions, extrapolated to the bound state pole, is derived from the 
Schr\"{o}dinger equation. It is shown that, unlike previous work, the result 
does not depend on the details of the potential through the corresponding Jost 
function but is given uniquely in terms of the binding energy. The 
generalisations to higher partial waves and one-dimensional scattering are 
given.
\end{abstract}
%
%
\newpage
\section{Introduction}

It is demonstrated in standard quantum mechanics texts \cite{Schiff} that when 
a scattering wave function is extrapolated to negative energy to the position of
a bound state then the result is proportional to the bound state wave function.
This is achieved through the divergence at this energy of the coefficient of the
exponentially decreasing function at large distances. 

Much less interest has been placed on the relative normalisations of the
scattering and bound state wave functions. Goldberger and Watson 
\cite{GW} and subsequently Joachain \cite{Joachain} have given a formal 
expression for this in terms of the 
Jost function and its derivative, which suggests that this quantity will in 
fact depend upon the form of the potential.

We have recently shown \cite{FW1}, using results from formal scattering theory,
that the continuation of an S-state scattering wave function with real boundary
conditions to a bound state pole actually only depends upon the bound state
wave function and binding energy. This result is counter-intuitive in that a
scattering wave is normalised by its asymptotic behaviour at large distances,
whereas a bound state is normalised by an integral condition. It is the purpose
of the present work to give a self-contained demonstration of this starting
from the Schr\"{o}dinger equation and our principal result,  eq.(\ref{2_25}),
is to be found in \S2. The generalisation of this to higher partial waves is
outlined in \S3, where the importance threshold kinematic factors in the
extrapolation to  the bound state is stressed.

The relationship to the Goldberger and Watson form is investigated in \S4,
where it is shown that, carried to its logical conclusion, their form would
give an answer identical to ours at the bound state position. Their
extrapolation is, however, less smooth as a function of energy and, as a
consequence, less useful in practice. 

The equivalent theorem for  one-dimensional scattering is derived in \S5 and
some discussion of the usefulness of the results given in the conclusions of
\S6. 
%
%
\newpage 
\section{Derivation of the Theorem}

We start by summarising the relevant results of potential scattering as given,
for example, in refs.\cite{Newton,Chadan}.

Consider a particle of mass $m$ moving with energy $E=k^2/2m$ and angular
momentum $\ell$ in a real spherically symmetric potential $V(r)$. In units
where $\hbar=1$, the radial Schr{\"o}dinger equation becomes
\begin{equation}
\frac{{\rm d}^2}{{\rm d}r^2}\,\psi_{\ell}(k,r)+
\left(k^2-\frac{\ell(\ell+1)}{r^2}\right)
\psi_{\ell}(k,r) = U(r)\, \psi_{\ell}(k,r)\,, \label{2_1}
\end{equation}
with $U(r)=2mV(r)$.

In order that the results of standard scattering theory apply, we assume that
the potential satisfies the integrability condition
\begin{equation}
\int_0^{\infty}{\rm d}r\,r\mid V(r)\mid \, <\infty \, .
\label{2_2}
\end{equation}
This requires that the potential be less singular than $r^{-2}$ at
short distances but decrease more rapidly than $r^{-2}$ at large
ones. For such a regular potential the bound state eigenvalues of
eq.(\ref{2_1}) are non-degenerate and finite in number \cite{Chadan}.

The generalisation to higher partial waves will be given in
\S3, and here we shall only consider S-wave solutions for which the index 
$\ell=0$ on the radial wave function will be suppressed.

The Jost solution $f(k,r)$ is determined by the asymptotic behaviour
\begin{equation}
f(k,r)\ \stackrel{r\to\infty}{\longrightarrow}\ \mbox{\rm e}^{ikr}\,.
\label{2_3}
\end{equation}
With this boundary condition the radial Schr{\"{o}}dinger equation
(\ref{2_1}) can be transformed into an integral equation and it 
is then straightforward to show by iteration that $f(k,r)$ is 
analytic in the half-plane Im$\{k\}>0$ and is continuous and bounded in 
Im$\{k\}\geq 0$ for all values of $r\geq 0$ \cite{Chadan}.

The value of this solution at $r=0$ is the Jost function
\begin{equation}
F(k) \equiv f(k,0)  \, ,\label{2_4}
\end{equation}
and, unless this vanishes, the full wave function resulting from the Jost
solution will be singular at the origin.

In the scattering region, $k$ is real and positive, whereas for
a bound state $k=i\alpha$, with $\alpha>0$ in order that the bound state be
normalisable. For any fixed value of $r$ the scattering wave function $f(k,r)$ 
can be analytically continued in $k$ to give the bound state wave function 
$f(i\alpha,r)$ normalised such that it behaves like $\mbox{\rm e}^{-\alpha r}$
at large distances.

A second independent solution of the Schr{\"{o}}dinger equation is $f(-k,r)$,
and it is easy to see from the integral form of the equation that for 
Im$\{k\}\geq 0$
\begin{equation}
\label{2_5}
f(-k^{\ast},r) = f(k,r) ^{\ast} \, ,
\end{equation}
from which it follows that for the Jost function
\begin{equation}
F(-k^{\ast}) = F(k) ^{\ast} \, . \label{2_6}
\end{equation}

For our purposes of continuing to the bound state pole, it is crucial to 
work with a \underline{real} function. A physical radial wave function $v(k,r)$
must vanish at the origin and the real linear combination with this property is
\begin{equation}
v(k,r) = \frac{1}{2ik}\left[ f(k,r)\,\mbox{\rm e}^{i\delta(k)}
 -f(-k,r)\,\mbox{\rm e}^{-i\delta(k)}\right] \,,
\label{2_7}
\end{equation}
with the phase shift $\delta(k)$ and corresponding S-matrix defined by
\begin{equation}
S(k)= \frac{F(-k)}{F(k)}= \mbox{\rm e}^{2i\delta(k)} \, .
\label{2_8}
\end{equation}

The function $v(k,r)$ is real, due to the symmetry property of the Jost 
function shown in eq.(\ref{2_6}), and behaves 
asymptotically like
\begin{equation}
v(k,r)\sim\frac{1}{k}\sin (kr+\delta(k)) \, .
\label{2_9}
\end{equation}

The bound state positions are determined by demanding that the wave function 
decrease at large $r$ and be finite at the origin, and this is achieved if
$F(i\alpha)=0$. The corresponding $S$-matrix has a pole at this position and
it is known that such poles are simple for a well-behaved potential satisfying
eq.(\ref{2_2}) \cite{Chadan}.

For positive real values of $k$ the norm of the $S$-matrix is, by 
eq.(\ref{2_6}), unity. We may parametrise it in the 
vicinity of an isolated bound state pole as
\begin{equation}
S(k) \equiv \mbox{\rm e}^{2i\delta(k)}=\frac{[N\,G(k)]^2}{\alpha+ik}\:, 
\label{2_10}
\end{equation}
where $N^2$ is the residue at the pole.
In the unitarised scattering length approximation $N^2=2\alpha$ and
$[G(k)]^2=(\alpha-ik)/2\alpha$, but in general $G(k)$ is an analytic function
of $k$ in the neighbourhood of $k=i\alpha$, with the condition that 
$G(i\alpha)=1$.

Comparing the radial S-wave Schr\"odinger equations 
for the bound and scattering state wave functions, we have
\begin{eqnarray}
u''(r) -\alpha^2u(r)&=&U(r)\,u(r) \nonumber \\
v''(k,r)+k^2v(k,r)&=&U(r)\,v(k,r) \, ,\label{2_11}
\end{eqnarray}
where $U(r)$ is a real function.

The boundary conditions at the origin are $u(0)=v(k,0)=0$, whereas the 
large-$r$ behaviour of the bound state wave function is taken to be
\begin{equation}
\label{2_12}
u(r)\ \ \stackrel{r\to\infty}{\longrightarrow}
\ \ \mbox{\rm e}^{-\alpha r} \, ,
\end{equation}
while that of $v(k,r)$ is given by eq.(\ref{2_9}). The reality of the potential,
combined with that of the boundary conditions, ensures that for real $k$ the 
functions $u(r)$ and $v(k,r)$ remain real for all values of $r$.

The wave function $v(k,r)$ can be analytically continued in $k$ to the bound
state at $k=i\alpha$, though care must be taken due to the singularity structure
of the factor $\mbox{\rm e}^{i\delta(k)}$. This is more complicated
than its square, the $S$-matrix, since it has a branch cut starting at the
position of the bound state.

It follows from eq.(\ref{2_10}) that at the bound state pole
\begin{equation}
\left[ \sqrt{2\alpha(\alpha^2+k^2)}\, v(k,r) \right]_{k=i\alpha}=
-\left[ \sqrt{\alpha+ik}\,\mbox{\rm e}^{i\delta(k)} \right]_{k=i\alpha}
f(i\alpha,r) =-N u(r) \, .             \label{2_13}
\end{equation}
It remains to be shown that the constant $N$ is determined uniquely by the
normalisation of the bound state wave function.

Manipulation of the radial Schr{\"{o}}dinger equations in  eq.(\ref{2_11})
leads to
\begin{equation}
\frac{{\rm d}}{{\rm d}r} \left[ u'(r)v(k,r)-u(r)v'(k,r) \right]
=(\alpha^2+k^2)u(r)v(k,r) \,, \label{2_14}
\end{equation}
where prime indicates derivative with respect to $r$. Since the functions
vanish at the origin, this can be integrated to give
\begin{equation}
 u'(r)v(k,r)-u(r)v'(k,r)
=(\alpha^2+k^2) \int_{0}^{r}{\rm d}r' u(r') v(k,r') \, . \label{2_15}
\end{equation}

Both sides of this equation vanish when $r\to\infty$, reflecting the 
orthogonality of the scattering and bound state wave functions. To avoid a 
$0=0$ statement, introduce the combination
\begin{equation}
\label{2_16}
w(k,r)\equiv 2ik\sqrt{\alpha+ik}\, v(k,r) \,,
\end{equation}
which, by eq.(\ref{2_13}), has the limit at the pole
\begin{equation}
\label{2_17}
w(i\alpha,r) = N\,u(r)\:.
\end{equation}

Differentiating the resulting
\begin{equation}
 u'(r)w(k,r)-u(r)w'(k,r)
=(\alpha^2+k^2) \int_{0}^{r}{\rm d}r' u(r') w(k,r')  \label{2_18}
\end{equation}
with respect to $k$ leads to
\begin{equation}
 u'(r)\dot{w}(k,r)-u(r)\dot{w}'(k,r)
=\int_{0}^{r}{\rm d}r' u(r') \left[ (\alpha^2+k^2) \dot{w}(k,r')
+2kw(k,r') \right]\, ,
\label{2_19}
\end{equation}
where derivatives with respect to $k$ have been indicated by dots.

After taking first the limit $k\to i\alpha$ and then $r\to \infty$,
the first term in the integrand vanishes because of the explicit
$(\alpha^2+k^2)$ factor, and so the right hand side of eq.(\ref{2_19})
becomes
\begin{equation}
\label{2_20}
-2iN\alpha\int_0^{\infty} dr\,u^2(r)\:.
\end{equation}

On the other hand, from the ansatz of eq.(\ref{2_10}), 
in the vicinity of the pole
\begin{eqnarray}
\nonumber
w(k,r) &= & N\,G(k)\,f(k,r)
-\frac{(\alpha+ik)}{N\,G(k)}\,f(-k,r)\, , \\[1ex]
\nonumber
\dot{w}(k,r) &= & 
N\,\dot{G}(k)\,f(k,r) +N\,G(k)\,\dot{f}(k,r)
-\frac{i}{N\,G(k)}\,f(-k,r)\\ 
&&+\frac{(\alpha+ik)}{N\,G(k)}\,\dot{f}(-k,r)
+\frac{(\alpha+ik)}{N\,G(k)^2}\,\dot{G}(k)\,f(-k,r)\, . 
\label{2_21}
\end{eqnarray}

Thus, since $G(i\alpha)=1$ at the pole,
\begin{eqnarray}
w(i\alpha,r) &=& N\,f(i\alpha,r)\ \stackrel{r\to\infty}{\longrightarrow}\
N\,\mbox{\rm e}^{-\alpha r}
\, , \label{2_22}\\[1ex]
\nonumber
 \dot{w}(i\alpha,r) &=& 
N\,\dot{G}(i\alpha)\,f(i\alpha,r) +N\,\dot{f}(i\alpha,r)
-\frac{i}{N}\,f(-i\alpha,r)
\\[1ex]
&\stackrel{r\to\infty}{\longrightarrow}&\ \
-\frac{i}{N}\,\mbox{\rm e}^{+\alpha r}\,.\label{2_23}
\end{eqnarray}

Since the bound state wave function decreases exponentially, the only surviving
contributions to the left hand side of eq.(\ref{2_19}) at large $r$
originate from the term proportional to $f(-i\alpha,r)$ in
eq.(\ref{2_23}), yielding a value of $-2i\alpha/N$.

Equating the two sides of eq.(\ref{2_19}), leads to the condition that
\begin{equation}
1=N^2 \int_{0}^{\infty}{\rm d}r\, u^2(r)  \, , \label{2_24}
\end{equation}
so that $N$ is indeed the normalisation constant of the bound state wave
function. 

Rewriting the result in terms of a normalised radial bound state
wave function $u_{\alpha}(r)=Nu(r)$, we see that the scattering wave function
$v(k,r)$ is related to $u_{\alpha}(r)$ at the pole $k=i\alpha$ through
\begin{equation}
\lim_{k\to i\alpha}
\left\{ \sqrt{2\alpha(\alpha^2+k^2)}\, v(k,r) \right\}= -
 u_{\alpha}(r) \:,              \label{2_25}
\end{equation}
and this is the principal result of this work.

It should be noted that, in contrast to the result given in ref.\cite{Joachain},
the constant of proportionality in this relation is
independent of the form of the potential and of the Jost function. Note also
that, since a particular bound state is not specified, 
eq.(\ref{2_25}) is valid at \underline{all} bound state 
poles.
%
%
\newpage 
\section{Higher partial waves}

For higher $\ell$-values the discussion given in \S2 can be repeated with only
minor changes. The Jost solutions of eq.(\ref{2_3}) are 
determined by the asymptotic behaviour \cite{Chadan}
\begin{equation}
\label{3_1}
f_{\ell}(\pm k,r)\ \stackrel{r\to\infty}{\longrightarrow}\ 
\mbox{\rm e}^{i\pi\ell/2}\mbox{\rm e}^{\pm ikr}\:,
\end{equation}
and the Jost functions defined as
\begin{equation}
\label{3_2}
F_{\ell}(\pm k)=\lim_{r\to 0}\left\{\frac{(\mp kr)^{\ell}}{(2\ell+1)!!}\,
f_{\ell}(\pm k,r)\right\}\:.
\end{equation}

The two Jost solutions are related for real $k$ by
\begin{equation}
\label{3_3}
f_{\ell}(-k,r)=(-1)^{\ell}\,\left[f_{\ell}(k,r)\right]^*\:,\\
\end{equation}
and similarly for the Jost functions.

The physical radial wave function $v_{\ell}(k,r)$ must vanish at the origin like
$r^{\ell +1}$, and the real linear combination with this property is
\begin{equation}
\label{3_4}
v_{\ell}(k,r) =\frac{1}{2ik}\left[(-1)^{\ell}f_{\ell}(k,r)
\mbox{\rm e}^{i\delta_{\ell}(k)}-
f_{\ell}(-k,r)\mbox{\rm e}^{-i\delta_{\ell}(k)}\right]\:,
\end{equation}
which for large values of $r$ tends to
\begin{equation}
\label{3_5}
v_{\ell}(k,r)\ \stackrel{r\to\infty}{\longrightarrow} \ 
\frac{1}{k}\sin(kr-\half\ell\pi+\delta_{\ell}(k))\:.
\end{equation}

In the vicinity of the bound state pole we write the S-matrix
\begin{equation}
\label{3_6}
S_{\ell}(k)\equiv \mbox{\rm e}^{2i\delta_{\ell}(k)} 
= \frac{F_{\ell}(-k)}{F_{\ell}(k)}= 
\frac{(-1)^{\ell}\left[N_{\ell}G_{\ell}(k)\right]^2}{\alpha+ik}\,,
\end{equation}
with $G_{\ell}(i\alpha)=1$ and $N_{\ell}$ real.

The residue has an $\ell$-dependent phase factor and we define the square root
through
\begin{equation}
\label{3_7}
\mbox{\rm e}^{i\delta_{\ell}(k)} = 
\mbox{\rm e}^{i\pi\ell/2}\frac{N_{\ell}G_{\ell}(k)}{\sqrt{\alpha+ik}}\,.
\end{equation}

The extra phase factor matches that in the definitions of eqs.(\ref{3_1}) and
(\ref{3_7}) such that, with $w_{\ell}(k,r)$ defined as in eq.(\ref{2_16}), the 
results of eqs.(\ref{2_17}) and (\ref{2_18}) remain unchanged with
$N_{\ell}$ the asymptotic normalisation constant for angular momentum $\ell$.
In particular, the bound state wave function with the asymptotic normalisation
of eq.(\ref{2_12}) is defined by
\begin{equation}
\label{3_8}
u_{\ell,\alpha}(r) = (-i)^{\ell} f_{\ell}(i\alpha,r)\:.
\end{equation}

In sum, our principal result of eq.(\ref{2_25}) remains true for arbitrary 
partial waves. However, in order to provide a \underline{smooth} extrapolation 
to the bound state pole it is preferable to take out the explicit threshold 
behaviour by dividing by a $k^{\ell}$ factor to give
\begin{equation}
\lim_{k\to i\alpha}
\left\{ \sqrt{2\alpha(\alpha^2+k^2)}\,\left(\frac{\alpha}{k}\right)^{\!\ell} 
v_{\ell}(k,r) \right\}= 
-(-i)^{\ell}\,u_{\ell,\alpha}(r) \:.              \label{3_9}
\end{equation}
Though of little importance in practical applications, this introduces an 
extra overall phase factor, whose origins may be traced to the definition of
the partial wave solution of eq.(\ref{3_4}).
%
%
\newpage
\section{Relation to the Goldberger and Watson Form}
\noindent

Using formal scattering theory arguments, Goldberger and Watson have derived 
an expression relating the normalisation of scattering and bound state wave 
functions \cite{GW} for the regular S-wave solution $\varphi(k,r)$, which is 
fixed by two real boundary conditions at the origin. The result, which depends 
upon the values of the Jost function and its derivative at the S-matrix pole, 
cannot be simplified, which means that the relation between the regular 
solution $\varphi(k,r)$ and the bound state wave function $u_{\alpha}(r)$ 
depends intrinsically upon the form of the potential.

The Goldberger-Watson form can however easily be recast to give a relation
for the physical scattering wave function $v(k,r)$ ({\it cf.} eq.(11.89) of 
ref.\cite{Joachain}).
\begin{equation}
\lim_{k\to i\alpha}
\left\{ \sqrt{\frac{4i\alpha^2F(k)}{\dot{F}(k)}}\, v(k,r) \right\}= -
 u_{\alpha}(r) \:.              \label{4_1}
\end{equation}

In the vicinity of an isolated zero the Jost function may be written as
\begin{equation}
F(k)=i\,{\cal C}(k)\,(k-i\alpha) \,,
\label{4_2}
\end{equation}
where  ${\cal C}(k)$ is regular and ${\cal C}(i\alpha)\neq0$. 

It then follows immediately that 
\begin{equation}
\lim_{k\to i\alpha}
\left\{\sqrt{\frac{4i\alpha^2{\cal C}(k)(k-i\alpha)}
{\dot{\cal C}(k)(k-i\alpha)+{\cal C}(k)}}\, v(k,r) \right\}
=\lim_{k\to i\alpha}
\left\{ \sqrt{4i\alpha^2(k-i\alpha)}\, v(k,r) \right\}
= - u_{\alpha}(r) \:.              
\label{4_3}
\end{equation}

At the pole this coincides with our form of eq.(\ref{2_25}),
as indeed it must, and so it is hard to understand why earlier developments 
\cite{Joachain} stopped just before the implementation of eq.(\ref{4_2}).

However for real values of $k$ our expression for the extrapolation function 
in eq.(\ref{2_25}) remains explicitly real, 
whereas that of eq.(\ref{4_3}) becomes complex. As a consequence 
the former provides a smoother extrapolation for the relationship
between two real quantities. 

To see the difference in practice, consider a one-term separable potential
with shape parameter $\beta$, for which the ratio of the scattering and bound
state wave functions at the origin is given by
\begin{equation}
\label{4_4}
R(k)\equiv \left\{\frac{v(k,r)}{u_{\alpha}(r)}\right\}_{r=0}
= \frac{1+2z}{\sqrt{2\alpha(k^2+\alpha^2)}\sqrt{1+z}}\:,
\end{equation}
where
\begin{equation}
\label{4_5}
z=\frac{k^2+\alpha^2}{4\beta(\alpha+\beta)}\:\cdot
\end{equation}
The correction terms to eq.(\ref{2_25}) are therefore small
at $r=0$ providing $k^2$ and $\alpha^2$ are small compared to $\beta^2$.

The extrapolation function for the Goldberger and Watson form may be easily
derived from the Jost function 
\begin{equation}
\label{4_6}
F(k)=\frac{(k-i\beta)[k+i(2\beta+\alpha)](k-i\alpha)}
{(k+i\beta)[k^2+(\alpha+\beta)^2+\beta^2]}\,.
\end{equation}
It is easily seen that this leads to large errors as soon as $k^2$ is 
comparable to $\alpha^2$ and this is due to the neglect of the influence of the
zero of the S-matrix at $k=-i\alpha$.
%
%
\newpage
\section{One-dimensional scattering}

One-dimensional scattering problems show certain features which are not
present in the more usual three-dimensional case \cite{Chadan}, but our proof
goes through here largely unaltered. 

Assuming the potential to be an even function of the variable $x$, for which 
$V(-x)=V(x)$, it is convenient to work with even and odd solutions of the
Schr\"{o}d\-inger  equation $v_{+}(k,x)$ and $v_{-}(k,x)$ which at the origin
satisfy the boundary  conditions
\begin{eqnarray}
\label{5_1}
v_{+}'(k,0) &=&0\:,\\
\label{5_2}
v_{-}(k,0)  &=& 0\:.
\end{eqnarray}

For large positive values of $x$ the functions behave like
\begin{eqnarray}
\label{5_3}
v_{+}(k,x)\ \ &\stackrel{x\to\infty}{\longrightarrow}&\ \ 
\phantom{-}\cos(kx+\delta_+(k))\:,\\
\label{5_4}
v_{-}(k,x)\ \ &\stackrel{x\to\infty}{\longrightarrow}&\ \ 
-\sin(kx+\delta_{-}(k))\:.
\end{eqnarray}

We define the normalisations of the corresponding symmetric and antisymmetric 
bound state wave functions through
\begin{equation}
\label{5_5}
\int_{-\infty}^{+\infty} dx \left[u_{+}^{\alpha}(x)\right]^2 =
\int_{-\infty}^{+\infty} dx \left[u_{-}^{\alpha}(x)\right]^2 =1\:.
\end{equation}

With the exception of an overall $-k$ factor in eq.(\ref{5_4}), the boundary
conditions on the antisymmetric function are identical to those of 
the S-wave three-dimensional problem of eq.(\ref{2_9}) combined with the
vanishing at the origin. Taking into account that the normalisation integral
extends down to $-\infty$ in eq.(\ref{5_5}), one can deduce the
one-dimensional extrapolation directly from the corresponding three-dimensional
case in eq.(\ref{2_25})
\begin{equation}
\lim_{k\to i\alpha}
\left\{\frac{1}{k}\,\sqrt{\alpha(\alpha^2+k^2)}\,v(k,x) \right\}= 
 u^{\alpha}(x) \:.              
\label{5_6}
\end{equation}

The presence of the $1/k$ factor in eq.(\ref{5_6}) does not upset the
smoothness of the extrapolation since this is compensated by the vanishing of 
$v_{-}(k,x)$ at $k=0$, which is a clear consequence of the antisymmetry of the 
function with respect to $kx$. 

For the symmetric case we must modify the arguments given in \S2. In the
vicinity of a simple pole of the S-matrix, we may write
\begin{equation}
\label{5_7}
S_{+}(k) \equiv \mbox{\rm e}^{2i\delta_{+}(k)} =
-\frac{2[N\,G(k)]^2}{\alpha+ik}\:,
\end{equation}
where $G(k)$ is analytic in the neighbourhood of the pole and is normalised
there to $G(i\alpha)=1$.

Now it is known that as $k\to 0$, $\delta_{+}(k)\to \pi/2$ unless there is a
zero-energy bound state \cite{Chadan}. This is one of the special
features of one-dimensional scattering, which causes the 
one-dimensional transmission coefficient to vanish at zero energy. When this
condition is imposed upon eq.(\ref{5_7}), it follows that
\begin{equation}
\label{5_8}
G(0) =\frac{1}{N}\sqrt{\frac{\alpha}{2}}\:\cdot
\end{equation}

Due to the symmetry of the problem, we need only consider the region $x\geq 0$,
and there the Jost solution is defined by its asymptotic behaviour
\begin{equation}
f(k,x)\ \stackrel{x\to\infty}{\longrightarrow}\  \mbox{\rm e}^{ikx}\,.
\label{5_9}
\end{equation}
This allows the even function to be written as
\begin{equation}
\label{5_10}
v(k,x)=\frac{1}{2}\left[f(k,x)\,\mbox{\rm e}^{i\delta_{+}(k)}
+f(-k,x)\,\mbox{\rm e}^{-i\delta_{+}(k)}\right] \,.
\end{equation}

After defining a non-normalised bound state wave function $u_{+}(x)$ with a
large-$x$ limit of e$^{-\alpha x}$, as in eq.(\ref{2_12}), all the subsequent
manipulations follow as in the three-dimensional proof. This is because
in the symmetric case the derivatives of the wave functions vanish at the
origin, and so it is still permissable to discard the left hand side of
eq.(\ref{2_15}) at $x=0$ to leave, as before,
\begin{equation}
\label{5_11}
 u_{+}'(x)v_{+}(k,x)-u_{+}(x)v_{+}'(k,x)
=(\alpha^2+k^2) \int_{0}^{x}{\rm d}x' u_{+}(x')v_{+}(k,x') \, . 
\end{equation}

It is then straightforward to show that the one-dimensional result of 
eq.(\ref{5_6}) is valid for both symmetric and antisymmetric 
wave functions.
%
%
\newpage
\vspace*{-2cm}
\section{Conclusions}

We have given a simple proof of the relation between the normalisations of the
bound and scattering wave functions at the bound state pole which is on a
par with the standard proof of the effective range expansion \cite{Schiff}.
Apart from being formally correct, the result is useful in practice because the
extrapolation function is smooth in $k^2$. As we saw in the case of the
separable potential in \S3, corrections to it only become important when $1/k$ 
is of the order of the range of the potential. It is therefore possible to use
the theorem to form approximations for scattering wave functions in terms of
that of the bound state. For the S-wave Paris neutron-proton triplet potential
\cite{Paris} such an approach is reliable out to radii of 1.7~fm for
centre-of-mass energies up to at least 20~MeV \cite{FW2}.

Already in 1952, Watson \cite{Watson} used such arguments to estimate the final
state interaction effects in $pp\to \pi^+(pn)$ at low excitation energies in
terms of the cross section for $pp\to \pi^+ d$. The relative scattering/bound
state normalisation was there established as an {\it approximation} using
effective range theory, though in practice it differs little from our result at
the pole. Providing the momentum transfer is large, as it is for pion
production, then the corresponding transition operator is sensitive to the 
short range part of the neutron-proton wave function where our theorem provides
a valid approximation. In addition to providing a useful description of 
$pp\to \pi^+(pn)$ data near threshold and at higher energies, it allows one to
understand quantitatively the final state interaction regions in 
$pn\to\eta(pn)$, $pd\to\pi^0(pd)$, and $dp\to p(pn)$ at high momentum transfers
\cite{FW1,FW2,BFW}.

The extension to higher partial waves described in \S3 is likely to be of less
significance since it is harder to investigate the effects of P-wave final state
interactions unless the S-wave is suppressed. Nevertheless the modifications to
our principal result of eq.(\ref{2_25}), engendered by threshold kinematic 
factors for $\ell\neq$~0, are of importance here.

Valuable discussions with B.~Karlsson are gratefully acknowledged. This work 
has been made possible by the continued financial support of the Swedish Royal 
Academy of Science and one of the authors (CW) would like to thank them and the
The Svedberg Laboratory for their generous hospitality.
%
%


\newpage
\baselineskip 4ex
\begin{thebibliography}{99}
\bibitem{Schiff} See for example L.I.~Schiff, {\it Quantum Mechanics}
(McGraw-Hill, N.Y., 1955).
\bibitem{GW} M.L.~Goldberger and K.M.~Watson, {\it Collision Theory}
(John Wiley \& Sons, N.Y., 1964).
\bibitem{Joachain} C.~Joachain, {\it Quantum Collision Theory} (North Holland,
Amsterdam, 1975).
\bibitem{FW1} G.F\"aldt and C.Wilkin, Nucl.Phys.~{\bf A604} (1996) 441.
\bibitem{Newton} R.G.Newton, {\it Scattering Theory of Waves and Particles}
(Springer-Verlag, N.Y., 1982).
\bibitem{Chadan} K.~Chadan and P.C.~Sabatier, {\it Inverse Problems in Quantum 
Scattering Theory} (Springer-Verlag, N.Y., 1989).
\bibitem{Paris} M.~Lacombe {\it et al.}, Phys.Rev.~{\bf C21} (1980) 861;
M.~Lacombe {\it et al.}, Phys.Lett. {B101} (1981) 139; B.~Loiseau, private
communication (1995).
\bibitem{FW2} G.F\"aldt and C.Wilkin, Phys.Lett.~{\bf B382} (1996) 209.
\bibitem{Watson} K.M.~Watson, Phys.Rev.~{\bf 88} (1952) 1163.
\bibitem{BFW} A.Boudard, G.F\"aldt and C.Wilkin, TSL/ISV-96-0142 (1996),
Phys.Lett.~{\bf B} (in press).
\end{thebibliography}
\end{document}